\documentclass[fleqn,usenatbib]{mnras}

\usepackage{mathptmx}

\usepackage[T1]{fontenc}

\usepackage{amssymb}	
\usepackage{graphicx}	
\usepackage{amsmath}	

\usepackage{hyperref}
\hypersetup{colorlinks=true,linkcolor=blue,citecolor=blue,filecolor=blue,urlcolor=blue}

\usepackage{siunitx}    

\usepackage[usenames, dvipsnames]{color}
\usepackage{soul}

\newcommand{\rninety}{$r_{90}$\,}

\usepackage{subcaption}

\title[X-ray properties of radio-AGN]{Fundamental differences in the X-ray accretion properties of low and high-excitation radio galaxies}

\author[R. Kondapally et al.]{R.~Kondapally,$^{1,2}$\thanks{E-mail: rohit.kondapally@durham.ac.uk}
T.~Holc,$^{3}$
J.~Aird,$^{3}$
K.~J.~Duncan,$^{3}$
P.~N.~Best,$^{3}$
D.~M.~Alexander,$^{1}$
S.~Das,$^{4}$
\newauthor{L.~K.~Morabito,$^{1,2}$
S.~Shenoy$^{4}$
and D.~J.~B.~Smith$^{4}$
}
\\
$^{1}$Centre for Extragalactic Astronomy, Department of Physics,
Durham University, South Road, Durham DH1 3LE, UK\\
$^{2}$Institute for Computational Cosmology, Department of Physics,
Durham University, South Road, Durham DH1 3LE, UK\\
$^{3}$Institute for Astronomy, University of Edinburgh, Royal Observatory, Blackford Hill, Edinburgh, EH9 3HJ, UK \\
$^{4}$Centre for Astrophysics Research, University of Hertfordshire, Hatfield AL10 9AB, UK\\
}

\date{Accepted XXX. Received YYY; in original form ZZZ}

\pubyear{2026}

\begin{document}
\label{firstpage}
\pagerange{\pageref{firstpage}--\pageref{lastpage}}
\maketitle

\begin{abstract}
We characterise the accretion rate properties of radio-detected AGN by combining deep radio and X-ray observations of the Bo\"{o}tes field. We used deep international LOFAR telescope observations to identify 2840 radio-AGN across $0.3 < z \leq 2$, divided into samples that are complete in radio luminosity. We further split this sample into four different classes: radio-quiet AGN (RQ-AGN), high-excitation radio galaxies (HERGs), and low-excitation radio galaxies (LERGs) hosted by star-forming (SF-LERGs) and quiescent galaxies (Q-LERGs). Through performing X-ray stacking, we determined the average X-ray luminosities, $L_{\rm{X, 2-10\,keV}}$ and average specific X-ray luminosities, $\lambda_{sL_{X}}$ (X-ray luminosity scaled by the stellar mass; a proxy for the accretion rate). We studied how these X-ray properties depend on radio luminosity, stellar mass, and redshift for each of the four AGN classes. We found that the LERGs, regardless of their star-formation activity, show significantly lower $L_{\rm{X, 2-10\,keV}}$ and $\lambda_{sL_{X}}$ than both HERGs and RQ-AGN across all redshifts. The average X-ray luminosities for the HERGs, RQ-AGN, and SF-LERGs typically increase with redshift, which may be associated with the increased cold gas fractions at earlier times, resulting in more enhanced black hole accretion. The average X-ray luminosities show weak-to-no correlation with radio luminosity at a given redshift, suggesting that the physical processes producing the X-ray and radio emission may not be coupled on spatial and/or temporal scales.
\end{abstract}

\begin{keywords}
galaxies: active -- galaxies: evolution -- galaxies: high-redshift -- radio-continuum: galaxies
\end{keywords}



\section{Introduction}\label{sec:intro}
Most massive galaxies in the local Universe are known to harbour a super massive black hole (SMBH) at their centre. Observations have shown correlations between the properties of SMBHs and their host galaxies, such as the tight relationship between the mass of the SMBH and that of the central galaxy bulge \citep{2000ApJ...539L...9F,2000ApJ...539L..13G}. Studies have also shown that the growth history of galaxies (traced by the cosmic star-formation rate density) is well correlated with the growth rate of the SMBHs, as traced by the black hole mass accretion rate density \citep[e.g.][]{2013ARA&A..51..511K,Aird2015}. Following these empirical relationships, it is now widely accepted that the build-up of galaxies and their black holes is likely connected across cosmic time. Accretion of matter from the surroundings grows these SMBHs, during which they are known as active galactic nuclei (AGN), releasing enormous amounts of radiation across the electromagnetic spectrum. This energetic output from the AGN can suppress star-formation and regulate subsequent growth of the galaxy \citep{2006MNRAS.370..645B,cattaneo2009_feedback_review,2012ARA&A..50..455F}; this process is known as AGN feedback and is thought to play a key role in the formation and evolution of galaxies \citep[see reviews by][]{Alexander2012,2014ARA&A..52..589H,Harrison2024,Alexander2025}. Feedback from radio-AGN, which emit jets of relativistic plasma observed at radio wavelength, are though to be particularly important in the formation of massive galaxies \citep[e.g.][]{2012MNRAS.421.1569B}.

Studies of radio-AGN in the nearby Universe have highlighted that AGN can be typically split into two different modes, commonly referred to as jet-mode and radiative-mode AGN \citep[e.g.][]{Allen2006,Hardcastle2007,2012MNRAS.421.1569B,2014ARA&A..52..589H,Harrison2024}. Jet-mode AGN display their characteristic powerful radio emission in the form of bi-polar jets launched by the SMBH, which is thought to undergo radiatively inefficient accretion ($\lesssim$ 1\% of the Eddington-scaled accretion rate). On the other hand, the radiative-mode AGN are typically associated with radiatively efficient accretion ($\gtrsim$ 1\% of the Eddington-scaled accretion rate), and can sometimes also display radio jets. The radiative-mode AGN are thought to be typically fuelled by cold gas, which leads to the formation of a geometrically thin, optically thick accretion disc \citep{1973A&A....24..337S}; as a result, the optical spectra of radiative-mode AGN show high-excitation emission lines. In contrast, the jet-mode AGN are typically fuelled by hot gas through an advection dominated accretion flow \citep{1994ApJ...428L..13N,1995ApJ...452..710N}; as a result, the jet-mode AGN lack a stable geometrically thin accretion disc and hence lack the high-excitation emission lines in their optical spectra. Owing to the nature of the emission lines in their optical spectra, the subset of radiative-mode AGN that display radio jets are commonly referred to as high-excitation radio galaxies (HERGs), whereas the jet-mode AGN are referred to low-excitation radio galaxies (LERGs).

In our current understanding, the HERGs and LERGs are two distinct populations of AGN, differentiated based on the Eddington-scaled accretion rate. In this scenario, the formation of a HERG requires an abundant supply of cold gas, which not only fuels the AGN but also star-formation within the galaxy. As a result, the host galaxies of HERGs tend to have lower stellar masses and higher star-formation rates (SFRs) compared to the host galaxies of the LERGs, which are typically massive quiescent galaxies with a hot gas halo that leads to radiatively inefficient accretion \citep[e.g.][]{2012MNRAS.421.1569B,Mingo2014,Janssen2012,Tadhunter2016,Ching2017b,Hardcastle2020}. However, recent work at both higher redshifts and fainter radio luminosities have found differences in the host galaxy properties of HERGs and LERGs compared to the nearby Universe \citep[e.g.][]{Kondapally2022,Whittam2022,Kondapally2025}. For example, \citet{Whittam2022} find a considerable overlap in the Eddington-scaled accretion rates for the HERGs and LERGs. Moreover, \citet{Kondapally2022} found that LERGs are predominantly hosted by star-forming galaxies at $z \gtrsim 1$, in contrast to their quiescent host galaxies at low redshift, casting doubt into how these LERGs may be fuelled in the early Universe. The Eddington-scaled accretion rates derived by \citet{Kondapally2025} using photometry and by \citet{Arnaudova2025,Arnaudova2026}, using spectroscopy, show a broad distribution for the LERGs, with a considerable fraction of SF-LERGs occupying accretion rates beyond $\sim 1\%$ of the Eddington rate. 

X-ray observations offer a complementary view of AGN activity, primarily arising from the inverse-Compton scattering of photons from the accretion disc by a hot corona, and as a result provide an alternative tracer of radiative accretion activity. Combining X-ray detections and non-detections of galaxies offers a powerful method of tracing the \textit{average} accretion properties of AGN in large galaxy samples \citep[e.g.][]{Hickox2007,Chen2013,Aird2017,Yang2017,Fornasini2018,Ito2022}. \citet{Delvecchio2018} performed such a study using X-ray stacking analysis for a radio-detected population using the VLA-COSMOS 3\,GHz Large survey, combining a sample of `radio-excess AGN' with deep \textit{Chandra} X-ray data. They were able to trace the average X-ray luminosities and accretion rates for the radio-AGN out to $z \sim 4$. They found that the average X-ray luminosities are weakly correlated with the radio-luminosity and that the radio-excess AGN trace more efficient AGN activity at higher redshifts; their analysis however did not study the LERGs and HERGs separately, which may show different accretion rate properties \citep[e.g.][]{2012MNRAS.421.1569B,Mingo2014}. In this paper, we combine data from the LOFAR Two Metre Sky Survey (LoTSS) Deep Fields \citep{Tasse2021,Kondapally2021,Duncan2021,Best2023} and the \textit{Chandra} Deep Wide-Field Survey (CDWFS; \citealt{Kenter2005,Masini2020}) to investigate the X-ray accretion rate properties of radio-detected AGN, through a stacking analysis. The high sensitivity and large survey area of LoTSS Deep Fields enables us to split our AGN sample into different classes and study how the X-ray properties evolve with redshift as a function of radio luminosity and stellar mass.







This paper is structured as follows. Sect.~\ref{sec:data} describes the radio, multi-wavelength and X-ray datasets we used, and outlines AGN identification and source classification. The stacking method used to obtain average X-ray properties is described in Section~\ref{sec:stacking}. Sect.~\ref{sec:xray_prop} presents our main results on the X-ray properties of different AGN classes. We present our discussion and comparison with results in literature in Sect.~\ref{sec:discussion}. Finally, we present our conclusions in Sect.~\ref{sec:conclusions}. Throughout this paper, a flat $\mathrm{\Lambda CDM}$ cosmology with $H_{\mathrm{0}} = 70~ \mathrm{km~s^{-1}~Mpc^{-1}}$, $\mathrm{\Omega_M = 0.3}$ and $\mathrm{\Omega_{\Lambda} = 0.7}$ is assumed, along with a radio spectral index $\alpha = -0.7$ (where $S_{\nu} \propto \nu^{\alpha}$).

\section{Data}\label{sec:data}
\subsection{Radio and associated multi-wavelength data}\label{sec:radio_data}
In this paper, we focus our analysis on the combination of radio, X-ray and other multi-wavelength data in the Bo\"{o}tes field. The parent AGN sample used in this study is constructed using international Low-Frequency Array (LOFAR) telescope High Band Antenna (HBA) data at $\sim$ 146\,MHz from the LOFAR Two-Metre Sky Survey (LoTSS) Deep Fields DR1 \citep{Tasse2021}. The LOFAR observations in Bo\"{o}tes reach an integration time of 96\,hours, yielding an rms sensitivity of $\sim 32\,\rm{\mu Jy\,beam^{-1}}$ at the centre of the field. Radio source detection was performed using Python Blob Detection and Source Finder (\textsc{PyBDSF}; \citealt{2015ascl.soft02007M}), extracting a catalogue down to the 30 per cent power of the primary beam.

The Bo\"{o}tes field consists of deep wide-area observations from the ultra-violet to far-infrared wavelengths which makes it ideal for studying the detailed properties of galaxies and AGN. The primary optical data in the \textit{B$_{\rm{w}}$}, \textit{R}, and \textit{I} filters come from the NOAO Deep Wide Field Survey (NDWFS; \citealt{1999ASPC..191..111J}). Near-infrared data in the J, H, and K\textsubscript{s} comes from \citet{gonzalez2010newfirmbootes} and mid-infrared data from the \textit{Spitzer} space telescope at 3.6, 4.5, 5.8 and 8.0\,$\mathrm{\mu}$m comes primarily from the Spitzer Deep Wide Field Survey (SDWFS; \citealt{eisenhardt2004iracshallow,2009ApJ...701..428A}). Additional far-infrared data covering 250, 350, and 500\,$\mu$m come from the \textit{Herschel} space telescope \citep{oliver2012hermes}. Using these observations, PSF matched multi-wavelength catalogues in the \textit{I}-band and 4.5\,$\mu$m band were generated by \citet{2007ApJ...654..858B,2008ApJ...682..937B}; these were optimally combined together by \citet{Kondapally2021}, which we use in this work. Photometric redshifts for the full multi-wavelength catalogue were generated by \citet{Duncan2021} using a combination of template fitting and machine learning methods suited for radio continuum surveys \citep{duncan2018photz_templates,duncan2018photz_ml}.

The multi-wavelength catalogues described above were used to correctly associate complex radio sources and identify their multi-wavelength counterparts, with this process described in detail by \citet{Kondapally2021}. In summary, radio source counterparts were identified by using a combination of the likelihood ratio method \citep{1977A&AS...28..211D,1992MNRAS.259..413S} for compact sources with well-defined positions and a visual classification scheme for extended or complex radio sources. This process resulted in a final radio catalogue of 19,179 sources with counterparts identified for over 97 per cent of these radio sources. For the vast majority of the radio sources, we use the photometric redshifts as determined by \citet{Duncan2021}. For $\sim$ 20 per cent of the sources in the field, good quality spectroscopic redshifts are available, which are used instead \citep[see][]{Duncan2021}.


\subsection{Source classification and identification of AGN}\label{sec:sclass_agn}
Faint radio-continuum surveys detect a mix of populations, including star-forming galaxies and AGN. \citet{Best2023} determined the host galaxy properties of the LOFAR detected sources in LoTSS-Deep DR1 and classified sources into different types of AGN using spectral energy distribution (SED) fitting. During this process, the redshift was fixed at the photometric redshift (or spectroscopic redshift, where available). In particular, every radio source was modelled using four different SED fitting codes: \textsc{bagpipes} \citep{Carnall2018}, \textsc{magphys} \citep{daCunha2008}, \textsc{agnfitter} \citep{CalistroRivera2017}, and \textsc{cigale} \citep{Boquien2019,Yang2020}, with the latter two codes being able to model emission from the AGN accretion disc and torus.

Using the output from the four SED fitting codes, \citet{Best2023} first identified the radiative-mode AGN. This was done by identifying sources where a considerable fraction of the total infrared luminosity emitted arose from the AGN components ($f_{\rm{AGN}}$) or by selecting sources where the $\chi^{2}_{\nu}$ goodness of fit for a given source indicated a better fit using codes that incorporated AGN models (\textsc{agnfitter} and \textsc{cigale}) compared to those that do not model AGN emission (\textsc{bagpipes} and \textsc{magphys}). For these radiative-mode AGN (also referred to as `SED AGN' by \citealt{Best2023}), key physical galaxy properties, such as the \textit{consensus} stellar mass and star-formation rate (SFR) were determined using the average of the results from \textsc{agnfitter} and \textsc{cigale}. In the Bo\"{o}tes field, X-ray observations from the X-Bo\"{o}tes survey \citep{Kenter2005} were used to identify additional `SED AGN'. In addition, for a small fraction of sources, optical spectroscopy of the radio sources suggested the presence of AGN, hence these were also classified as `SED AGN'.

\begin{figure}
    \centering
    \includegraphics[width=\columnwidth]{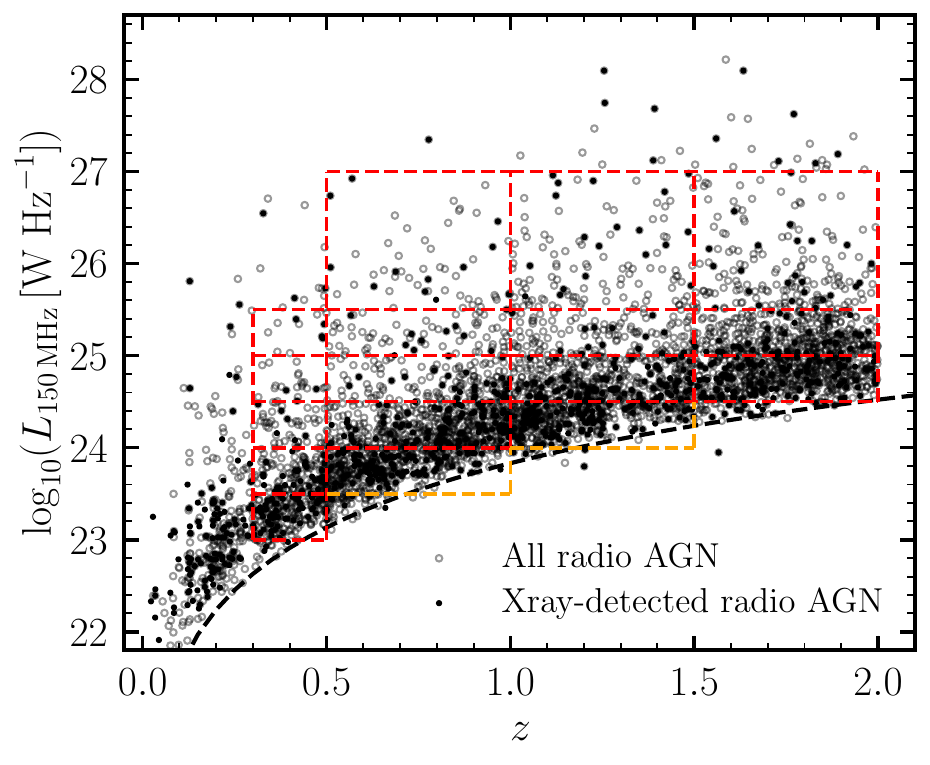}
    \caption{\label{fig:l150_z_bins}Distribution of sources across the radio-luminosity versus redshift plane in the Bo\"{o}tes field. The radio-detected AGN (open black circles) and the subset of these that are X-ray detected (filled back circles) are shown. The black dashed line shows the 5$\sigma$ radio luminosity limit in the Bo\"{o}tes field. Above this limit, the sample is divided into 15 complete bins in radio luminosity (shown in red), as well as a further 2 incomplete bins (shown in orange). We focus our analysis in Sect.~\ref{sec:xray_prop} of the X-ray properties as a function of radio luminosity using these bins.}
\end{figure}

There exists a well-studied correlation between the radio luminosity and SFR of star-forming galaxies, which can be used to select `radio-loud' AGN as sources with considerable excess radio emission above that expected based on this correlation \citep[e.g.][]{CalistroRivera2016,smolcic2017_vla_3ghz_counterparts,Delvecchio2017,gurkan2018lofar_sfr,Smith2021,Das2024}; these are hereafter referred to as `radio-excess AGN'. For the LoTSS-Deep DR1 sample, \citet{Best2023} selected radio-excess AGN as sources whose radio luminosity was 3$\sigma$ above that expected from star-formation processes alone.

The combination of the `radiative-mode AGN' and `radio-excess AGN' criteria were used by \citet{Best2023} to classify different types of radio-detected AGN as follows. Sources satisfying both radio-excess and radiative-mode AGN criteria were classified as HERGs; those identified as radio-excess but not radiative-mode were classed as LERGs. Finally, sources which were identified as radiative-mode but not radio-excess were classed as radio-quiet AGN (RQ-AGN). Sources identified using this classification scheme have been studied extensively and validated through various measurements, including luminosity functions and host-galaxy properties \citep[e.g.][]{Kondapally2022,Bonato2021,Cochrane2023}. Hence, we use this classification scheme to identify different radio-detected AGN in this paper. \citet{Kondapally2022,Kondapally2025} found that there is a considerable population of LERGs hosted by both quiescent and star-forming galaxies, which appear to show differences in their host-galaxy properties and possibly their accretion rate properties. Therefore, in this paper, we split the LERG population into those hosted by quiescent and star-forming galaxies (SFGs), and study the X-ray properties of the two populations separately. We select the quiescent host galaxies of LERGs using $\mathrm{sSFR} ~<~ 0.2/t_\mathrm{H}(z)$, where $\mathrm{sSFR} \equiv \rm{SFR}/M_{\star}$ is the specific star-formation rate, and $t_\mathrm{H}(z)$ is the age of the Universe at redshift $z$ \citep[e.g.][]{Pacifici2016}. In total, across the chosen redshift range for this study ($0.3 < z \leq 2$), our final sample contains 3496 radio-detected AGN. Of these, we have 334 HERGs, 1128 RQ-AGN, 844 LERGs hosted in quiescent galaxies (Q-LERGs hereafter), and 1190 LERGs hosted in star-forming galaxies (SF-LERGs hereafter). 

One of the aims of our paper is to study the X-ray properties of the radio sources as a function of radio luminosity. To avoid biasing our results due to the flux-limited nature of our survey, we divide the AGN sample into bins that are complete in radio luminosity across a given redshift interval. Fig.~\ref{fig:l150_z_bins} shows radio luminosity against redshift for all radio-detected AGN (open black circles) and their X-ray detected subset (filled black circles; see Sect.~\ref{sec:xray_data}). The black dashed line indicates the $5\sigma$ luminosity limit of radio data. We construct 15 bins complete in radio luminosity (red dashed lines in Fig.~\ref{fig:l150_z_bins}), as well as 2 additional incomplete bins (orange dashed lines); we include these in our analysis as they are close to the 5$\sigma$ limit but clearly distinguish them from luminosity-complete bins. In summary, after applying this radio luminosity selection, across $0.3 < z \leq 2$, our final sample includes 2840 radio-detected AGN, of which we have 298 HERGs, 839 RQ-AGN, 659 Q-LERGs, and 1044 SF-LERGs.


\subsection{X-ray data}\label{sec:xray_data}
X-ray data in the Bo\"{o}tes field was obtained from the \textit{Chandra} space telescope, as part of the \textit{Chandra} Deep Wide-field Survey (CDWFS) data release \citep{Masini2020}. The \textit{Chandra} observations of the Bo\"{o}tes field were carried out between 2003 and 2018, which includes the XBo\"{o}tes survey \citep{Kenter2005, murray2005_xbootes} and the Cycle 18 Large Program (PI: Hickox). The observations cover 281 pointings across $\sim$ 9\,deg$^{2}$ with a total exposure time of 3.4\,Ms. \citet{Masini2020} performed the data reduction and calibration of these \textit{Chandra} observations in a homogenous manner and then performed source extraction and characterisation in three X-ray bands: the broad (0.5--7 keV), soft (0.5--2 keV) and hard (2--7 keV) frequency bands. The details of the data reduction and source identification are provided by \citet{Masini2020}. In summary, 6891 X-ray sources were identified based on a detection threshold corresponding to a very low Poisson probability of the detected counts being the result of a background fluctuation. The multi-wavelength counterparts to the X-ray detected sources, including photometric redshifts, were identified via a cross-match to the same multi-wavelength and photometric redshift catalogue as the radio sample \citep{Kondapally2021,Duncan2021}. Hence, starting with the radio-detected sample defined in Sect.~\ref{sec:sclass_agn}, we directly match to the X-ray counterpart positions to identify the X-ray detected radio-AGN. Overall, we find 133 of the HERGs are X-ray detected (44.6\% of the parent radio-detected HERG population), and similarly, 328 RQ-AGN (39.1\%), 26 Q-LERGs (3.9\%), and 43 SF-LERGs (4.1\%) are X-ray detected.




\section{X-ray stacking of radio-detected sources}\label{sec:stacking}
The vast majority of the radio-detected AGN are not X-ray detected, as noted in Sect.~\ref{sec:xray_data}. Hence, we perform an X-ray stacking analysis to study the average X-ray properties of different AGN types, incorporating X-ray information for the non-detections. In this paper, we study the X-ray properties for each of the HERGs, RQ-AGN, Q-LERGs, and SF-LERGs across four redshift bins, $0.3 < z \leq 0.5$, $0.5 < z \leq 1$, $1 <z \leq 1.5$, and $1.5 < z \leq 2$. As detailed in Sect.~\ref{sec:xray_prop}, in each redshift bin, we study the average X-ray properties of these AGN as a function of the 150\,MHz radio luminosity, stellar mass, and redshift. Therefore, in practice, in each redshift bin, the X-ray stacking analysis is performed by splitting the sample into bins of radio luminosity or stellar mass, as appropriate, for each of the four AGN categories.

Our aim is to study the \textit{average} X-ray properties of the different types of radio-AGN, hence we chose to combine both the X-ray detections and the X-ray non-detections in our stacks \citep[e.g.][]{Delvecchio2018}. This is particularly important for the HERGs and RQ-AGN, where a significant fraction of the sources is X-ray detected (see Sect.~\ref{sec:xray_data}); excluding these sources would bias our results and subsequent interpretation of the population as a whole. To perform the X-ray stacking analysis, we use the properties of the individual detections from \citet{Masini2020} where possible. In this analysis we focus primarily on the \textit{Chandra} hard band (2-7\,keV). The average X-ray flux, $\langle F_{X} \rangle$ was calculated following a similar process as that described by \citet{Chen2013} and \citet{Yang2017}, as follows:
\begin{equation}\label{eq:xray_stack_flux}
    \langle F_{X} \rangle = \frac{\sum_{j}^{N_{\rm{detected}}}F_{j,\rm{detected}} + F_{\rm{undetected}}}{N_{\rm{detected}} + N_{\rm{undetected}}},
\end{equation}
where $F_{j,\rm{detected}}$ is the flux of each detected source, $N_{\rm{detected}}$ is the number of X-ray detected sources, $N_{\rm{undetected}}$ is the number of X-ray undetected sources in the stack, and $F_{\rm{undetected}}$ is the total stacked flux from the X-ray non-detections. 

We calculated $F_{j,\rm{detected}}$ using the measured count rates in the catalogue of \citet{Masini2020}, which were then converted to fluxes using the energy conversion factor (ECF), which accounts for the varying effective area and instrument energy response over the long time-span on the observations. For this, we assumed a power-law spectrum with a photon index $\Gamma = 1.8$ \citep[e.g.][]{Ricci2017} and Galactic absorption $N_{\rm{\rm{H, Gal}}} = 1.04 \times 10^{20}\,\rm{cm^{-2}}$ \citep{Kalberla2005}. The total stacked flux, $F_{\rm{undetected}}$ was calculated as
\begin{equation}\label{eq:nondet_stack}
F_{\rm{undetected}} = \sum_{i}^{N_{\rm{undetected}}}\frac{1.1 \times \left(T_{i} - B_{i}\right)}{t_{\rm{exp,i}}} \mathbb{C}_{i},
\end{equation}
where, $B_{i}$ and $T_{i}$ are the background and the total counts, measured using a circular aperture of radius \rninety, which is commonly used to approximate the size of the point-spread function (PSF) across the field of view, $t_{\rm{exp,i}}$ is the average exposure time, and $\mathbb{C}_{i}$ is the average exposure-weighted ECF. 
During the stacking process, we exclude any sources from the stack if they are within $2 \times$ \rninety of an X-ray detected source; this avoids the contamination of the stacked signal from a nearby X-ray bright source.

To derive the final stacked fluxes and associated uncertainties, we performed a bootstrap resampling analysis of the input source list using Eq.~\ref{eq:xray_stack_flux}, with this process repeated 10,000 times (e.g. in each redshift, radio luminosity, or stellar mass bin; see Sect.~\ref{sec:xray_prop}). We take the median of the resulting distribution as our best-estimate of the stacked X-ray flux within a given bin, with the uncertainties estimated using the 16th and 84th percentiles of this distribution from bootstrapping. The average stacked 2-7\,keV fluxes are then scaled to the 2-10\,keV fluxes and then converted to rest-frame 2-10\,keV X-ray luminosities $L_{X,\rm{2-10\,keV}}$, using a \textit{k}-correction of $(1 + z_{\rm{median}})^{\Gamma-2}$ (with $\Gamma=1.8$), where $z_{\rm{median}}$ is the median redshift of a given sub-sample in a given bin. 


\subsection{Correction for contribution from XRBs}\label{sec:xrb_corr}
The X-ray emission from a galaxy primarily comes from AGN and star-formation processes. For the latter, the source of the X-ray emission in our stacks could have contributions from both the high-mass X-ray binaries (XRBs), which correlate well with the SFR, and from the low-mass XRBs which are well correlated with the stellar mass \citep[e.g.][]{Lehmer2010,Mineo2014,Lehmer2016,Aird2017,Fornasini2018}.

To model the contribution from the XRB population to our stacked X-ray luminosities, we used the best-fit relation determined by \citet{Lehmer2016}, given as
\begin{equation}\label{eq:xrb_corr}
\left(\frac{L_{\rm{XRB, 2-10\,keV}}}{\rm{erg\,s^{-1}}}\right) = 29.37(1+z)^{2.03}M_{\star} + 39.28(1+z)^{1.31}\rm{SFR},
\end{equation}
where $M_{\star}$ and $\rm{SFR}$ are the stellar mass. We determined the typical X-ray emission expected from XRBs by using the median stellar mass and SFR for each bin used in Sect.~\ref{sec:xray_prop}, and subtracted this from the 2-10\,keV  stacked X-ray luminosity determined above. The resulting AGN-related X-ray luminosity is used in the rest of the analysis.

\subsection{Assessing the impact of intrinsic obscuration}\label{sec:obs_corr}
\begin{figure}
    \centering
    \includegraphics[width=\columnwidth]{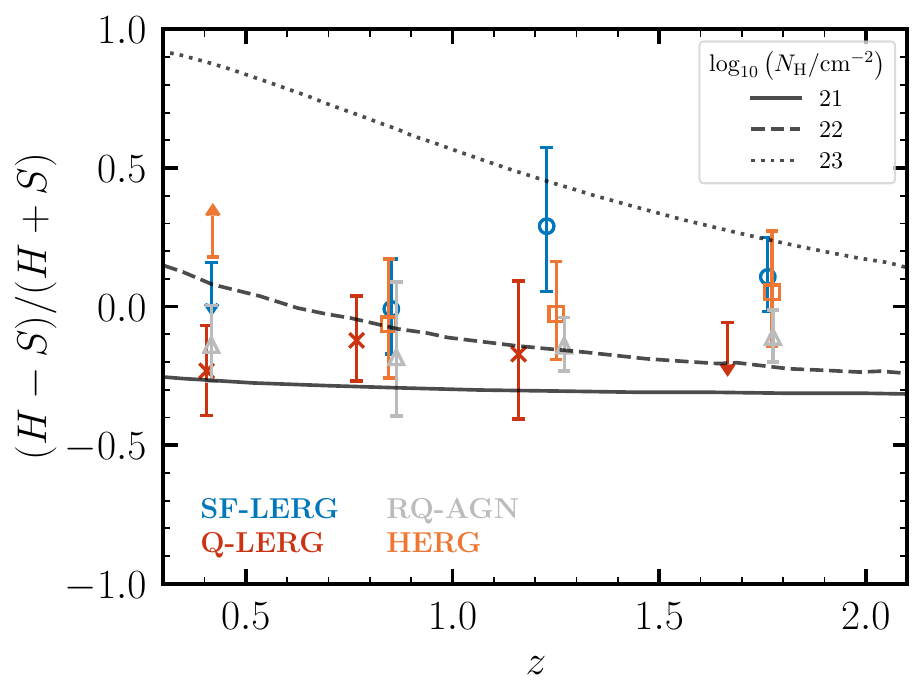}
    \caption{\label{fig:HR_z}Hardness ratios, $\rm{HR} = (H - S)/(H + S)$, as a function of redshift for each of the HERGs (orange), RQ-AGN (grey), SF-LERGs (blue), and the Q-LERGs (red). The HRs are determined by performing the X-ray stacking analysis in the \textit{Chandra} hard and soft bands. The curves correspond to the expected HRs for a power law spectra with $\Gamma = 1.8$, a Galactic absorption of $N_{\rm{\rm{H, Gal}}} = 1.04 \times 10^{20}\,\rm{cm^{-2}}$ \citep{Kalberla2005}, and then absorption through varying intrinsic column densities, $\log_{10}(N_{\rm{H}}/cm^{-2}) = 21, 22,$ and $23$,  shown by the solid, dashed, and dotted lines, respectively.}
\end{figure}

It is possible that the derived stacked X-ray fluxes may be underestimated due to intrinsic absorption. To estimate the level of obscuration and the impact that this could have on our results, we estimate the hardness ratios, $\rm{HR} = (H-S)/(H+S)$, where $H$ and $S$ are the hard (2-7\,keV) and soft (0.5-2\,keV) net counts, respectively. Fig.~\ref{fig:HR_z} shows the HRs for the four AGN classes as a function of redshift. The HRs are derived by performing the stacking analysis in both the hard and soft \textit{Chandra} bands, and are performed separately for the HERGs (orange squares), RQ-AGN (grey triangles), Q-LERGs (red crosses), and SF-LERGs (blue circles). For the bins with stacked soft-band SNR $< 2$, the 2$\sigma$ lower limits are instead shown. Also shown on the plot are model HRs expected for intrinsic column densities, $\log_{10}(N_{\rm{H}}/\rm{cm^{-2}}) = $ 21, 22, 23. The model HR curves are taken from the values computed by \citet{Masini2020} for our X-ray dataset. In summary, the HRs were computed by using \textsc{xspec} to model power law spectra (with $\Gamma = 1.8$), modified to include Galactic absorption by $N_{\rm{\rm{H, Gal}}} = 1.04 \times 10^{20}\,\rm{cm^{-2}}$ \citep{Kalberla2005}, and then absorbed by the above column densities.

\begin{figure*}
    \centering
    \includegraphics[width=\textwidth]{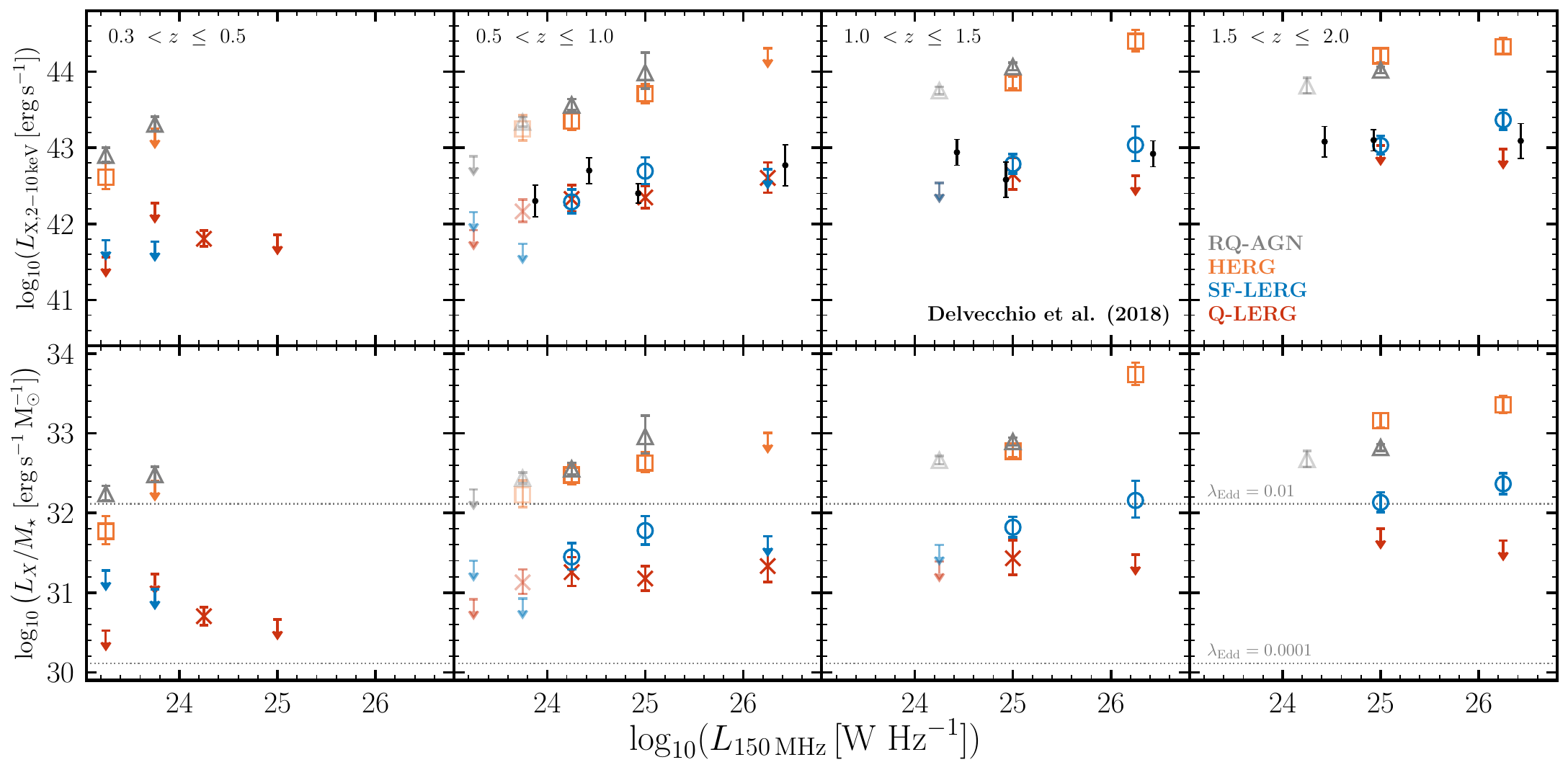}
    \caption{\label{fig:lx_bhar_l150}Stacked X-ray luminosities, $L_{X,\rm{2-10\,keV}}$ (\textit{top row}) and specific X-ray luminosities, $\lambda_{sL_{X}}$ (\textit{bottom row}) as a function of radio luminosity for radio-detected AGN across $0.3 < z \leq 2$. The stacked X-ray properties were calculated using the \textit{Chandra} hard band for each of the HERGs (orange), RQ-AGN (grey), SF-LERGs (blue), and the Q-LERGs (red). For bins where the stacked X-ray flux has a SNR < 2, we instead show the 2$\sigma$ upper limits. The stacked 2-10\,keV X-ray luminosities are calculated by incorporating both X-ray detections and non-detections, and corrected for contribution from XRBs. The X-ray luminosities are scaled by the stellar mass to derive the specific X-ray luminositiess (see text). The uncertainties are generated using a bootstrap resampling process. The data points shown in lighter colours indicate bins where the flux-limited LOFAR sample is incomplete over a given redshift range (as shown in Fig.~\ref{fig:l150_z_bins}). The grey lines in the bottom row show the specific X-ray luminosities corresponding to Eddington-scaled accretion rates of 1\% and 0.01\% (see Sect.~\ref{sec:discussion}). The results for the overall radio-excess AGN population from \citet{Delvecchio2018} are in good agreement with the stacked X-ray luminosities of LERGs, and lie consistently below the HERGs. We find that LERGs (regardless of their star-formation activity) have significantly lower average X-ray luminosities (and hence specific X-ray luminosities) compared to the HERGs and RQ-AGN. The radio-detected AGN typically show a weak dependence of $L_{X,\rm{2-10\,keV}}$ on radio luminosity at a given redshift bin.}
\end{figure*}

We find that on average, for a given redshift bin, the HRs of the different AGN classes are consistent with each other within the uncertainties. The HRs derived for the AGN samples suggest moderate levels of obscuration for all classes $\log_{10}(N_{\rm{H}}/\rm{cm^{-2}}) \sim 22 - 23$. For such typical $N_{\rm{H}}$ values, the absorption-corrected 2-7\,keV fluxes would be a factor of $\sim$ 5-10 per cent higher; this would correspond to a difference of $\lesssim 0.05\,\rm{dex}$ in log X-ray luminosity space. A key aim of this paper is to study the differences in the X-ray properties \textit{between} the different AGN classes. Hence, given the similar levels of HR-based column densities across the different AGN populations at a given redshift, the large uncertainties in determining HRs for stacked samples, and the relatively small correction factors, we decide to not apply absorption corrections to our stacked X-ray luminosities studies in Sect.~\ref{sec:xray_prop}; such corrections will have a negligible impact on the qualitative trends and results discussed in this paper.





\section{X-ray properties of radio-detected AGN}\label{sec:xray_prop}
In this Section, we use the process described in Sect.~\ref{sec:stacking} to study the average (stacked) X-ray luminosities of the radio-detected AGN.

The X-ray AGN luminosity can be used to trace the black hole accretion rates (BHAR) by adopting appropriate conversions to AGN bolometric luminosities and accretion efficiency \citep[e.g.][]{Vasudevan2007,Lusso2012,Duras2020}, and then scaling this to the Eddington luminosity to determine the Eddington-scaled accretion rates ($\lambda_{\rm{Edd}}$). However deriving appropriate bolometric corrections and estimating black hole masses from stellar masses can introduce additional uncertainties and scatter in the accretion rates, complicating the interpretation of the results. Hence, in this work, we instead define the average specific X-ray luminosity, $\lambda_{sL_{X}}$:
\begin{equation}\label{eq:sbhar}
    \lambda_{sL_{X}} = \frac{\langle L_{X,\rm{2-10\,keV}}\rangle}{M_{\star}},
\end{equation}
where $M_{\star}$ is the median stellar mass for a given sub-sample in a given bin. Defined in this manner, $\lambda_{sL_{X}}$ (with units of $\rm{erg\,s^{-1}\,M_{\odot}^{-1}}$) is analogous to the specific black hole accretion rate quantity defined by \citet{Aird2012} and \citet{Delvecchio2018}, and acts as a proxy for the accretion rate while not suffering from the above uncertainties or assumptions in deriving bolometric luminosities or Eddington-scaled accretion rates. Moreover, the X-ray bolometric corrections derived in the literature are applicable to the radiative-mode AGN population (i.e. the HERGs and RQ-AGN), however, they are likely not suitable for the AGN which are not expected to show significant radiative accretion (i.e. the LERGs). Hence, Eq.~\ref{eq:sbhar} also enables us to compare the X-ray properties of LERGs with HERGs and RQ-AGN in a consistent manner.

In Fig.~\ref{fig:lx_bhar_l150}, we show the average $L_{\rm{X, 2-10\,keV}}$ (top panel), and the average $\lambda_{sL_{X}}$ (bottom panel) as a function of the radio luminosity ($L_{\rm{150\,MHz}}$) for four different redshift bins (in four panels from left to right), $0.3 < z \leq 0.5$, $0.5, < z \leq 1$, $1 < z \leq 1.5$, $1.5 < z \leq 2$. We computed this using the methodology outlined in Sect.~\ref{sec:stacking} for the Q-LERGs (red crosses), SF-LERGs (blue circles), HERGs (orange squares), and RQ-AGN (grey triangles). The stacked X-ray luminosities ($L_{\rm{X, 2-10\,keV}}$) were corrected for the contribution from XRBs following equation~\ref{eq:xrb_corr}, using the median stellar mass and SFR for each source type within a given redshift and radio luminosity bin, as discussed in Sect.~\ref{sec:xrb_corr}. The $1\sigma$ uncertainties are calculated based on the 16th and 84th percentiles of this distribution. Bins with less than 10 sources are excluded as their stacked fluxes derived from a small number of sources may not be robust. In bins where the hard band stacked flux has a SNR $<$ 2, we display the $2\sigma$ upper limits for the X-ray luminosity and the corresponding specific X-ray luminosity.

We also compare our results to \citet{Delvecchio2018}, who utilised the VLA-COSMOS 3\,GHz Large Survey \citep{smolcic2017_vla_cosmos_3G,smolcic2017_vla_3ghz_counterparts} and combined this with the \textit{Chandra} COSMOS-Legacy survey \citep{Civano2016,Marchesi2016} to investigate the average X-ray properties of radio-excess AGN via X-ray stacking. We compare their stacked X-ray luminosities (scaled to 2-10\,keV; black points) with our results in Fig.~\ref{fig:lx_bhar_l150}. We convert the 1.4\,GHz radio luminosities from \citet{Delvecchio2018} to 150\,MHz using a standard spectral index $\alpha = -0.7$. We note that the redshift bins used by \citeauthor{Delvecchio2018} (\citeyear{Delvecchio2018}; $0.6 < z \leq 1$, $1 < z \leq 1.4$, and $1.4 < z \leq 1.8$) are slightly different to the ones used in this study. We therefore overlay their results in the closest redshift bin for comparison, but note than re-calculating our results to match their redshift bins does not qualitatively affect the overall results. 

We find that the results from \citet{Delvecchio2018} agree well with our stacked X-ray luminosities of LERGs, and lie significantly below our measurements for the HERGs. Although the exact source classification scheme adopted in \citet{Delvecchio2018} and our work are different, in principle, their `radio-excess AGN' sample will contain a mixture of LERGs and HERGs. Indeed, their radio-excess AGN sample, which is based on the parent catalogue by \citet{smolcic2017_vla_3ghz_counterparts}, contains a majority of LERG-like AGN (`MLAGN' in their terminology), with HERG-like AGN forming $\sim$ 25 per cent of their radio-excess population. Hence, their stacked measurements will be averaging over both the HERGs and the more numerous LERGs in their sample. LERGs are also the numerically dominant radio-excess AGN population in our sample, and hence it is not surprising that their average stacked X-ray luminosities agree more with that of our LERGs. We also note that the LoTSS-Deep data in Bo\"{o}tes covers a larger area than the VLA observations of COSMOS, and hence we expect to be more sensitive to the HERGs which tend to be rarer, higher radio luminosity AGN. These results highlight the importance of studying the properties of LERGs and HERGs separately.

\begin{figure*}
    \centering
    \includegraphics[width=\textwidth]{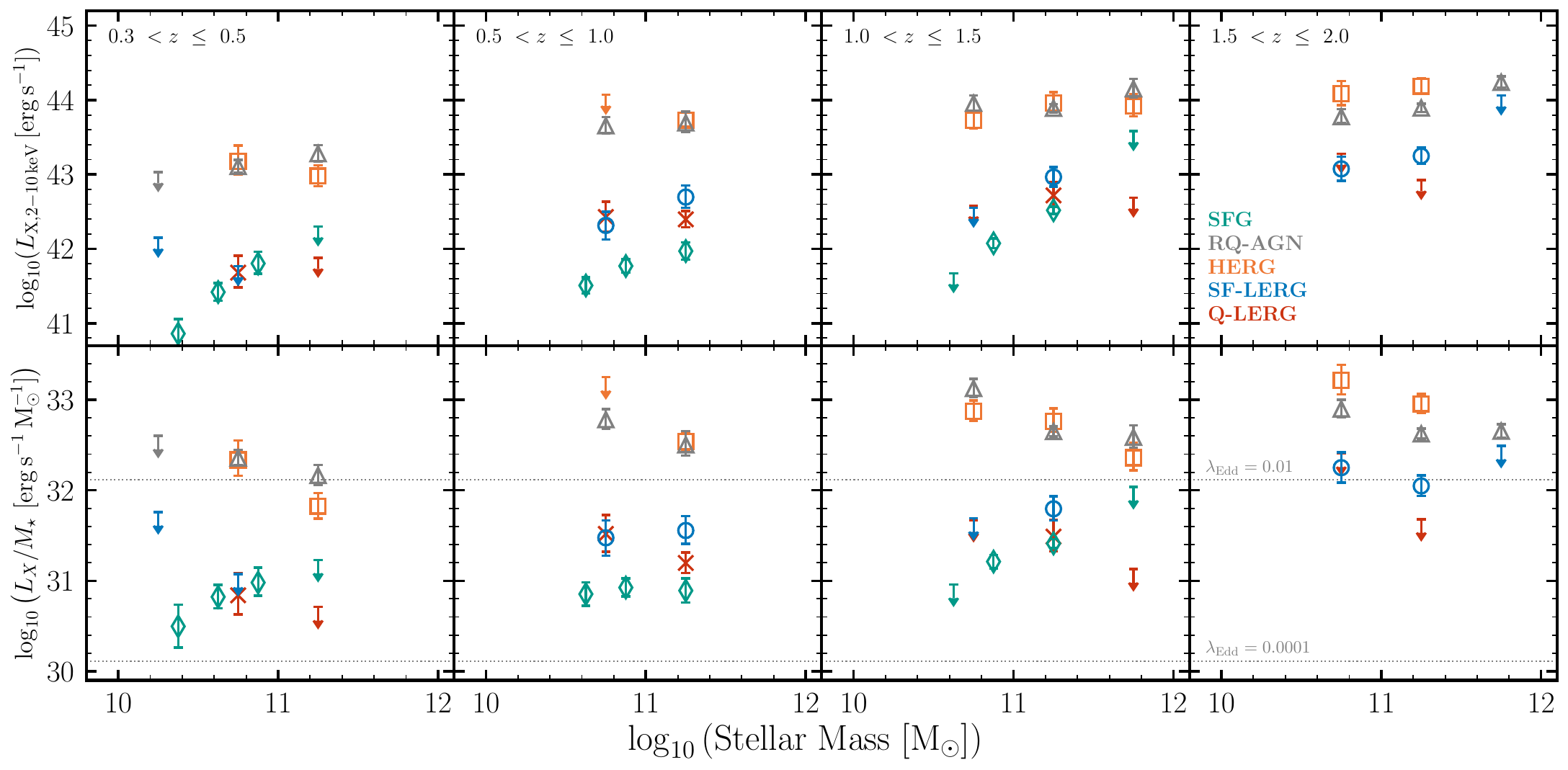}
    \caption{\label{fig:lx_bhar_mass}Stacked X-ray luminosities (\textit{top row}) and specific X-ray luminosities (\textit{bottom row}) as a function of stellar mass for radio-detected AGN across $0.3 < z \leq 2$. The markers and colours are the same as in Fig.~\ref{fig:lx_bhar_l150}. For bins where the stacked X-ray flux has a SNR < 2, we instead show the 2$\sigma$ upper limits. The uncertainties are generated using a bootstrap resampling process. We apply 90\% stellar mass completeness limits derived by \citet{Duncan2021}, by removing sources below this limit at a given redshift interval, to minimise biases (see text). We also compute the average $L_{X,\rm{2-10\,keV}}$ for a mid-infrared flux-selected sample of SFGs in the Bo\"{o}tes field (green diamonds) for comparison with the radio-detected AGN samples. For the radio-detected AGN, we find a weak correlation in the average X-ray luminosity with stellar mass at a given redshift bin. On average, the typical $L_{X,\rm{2-10\,keV}}$ and $\lambda_{sL_{X}}$ increase with redshift, indicating that the radio-detected AGN are typically accreting more efficient at earlier times for a fixed stellar mass.}
\end{figure*}

We note that it is possible that any potential differences in the typical stellar masses of different AGN classes may affect the observed results. We compared the median stellar masses across different AGN classes at $z > 0.5$ and found that these are broadly consistent with each other (within $\sim$ 0.3\,dex). The effects of this can be seen indirectly when comparing the relative shifts of the data points between the top and bottom panels of Fig.~\ref{fig:lx_bhar_l150}, where the X-ray luminosities are scaled by the median stellar mass. 
To study the dependence of the average $L_{X,\rm{2-10\,keV}}$ and $\lambda_{sL_{X}}$ on stellar mass more directly, we repeat our X-ray stacking analysis for the AGN samples in bins of stellar mass across redshift; the results are shown in Fig.~\ref{fig:lx_bhar_mass}. For this analysis, we also applied the 90\% stellar mass completeness limits derived by \citet{Duncan2021} by only including sources in the stack that could be detected across an entire redshift bin; this minimises the effect of observational biases on the stacked results. In addition to the radio-detected AGN samples, we also compute the average X-ray properties for a mid-infrared (3.6\,$\mu$m) flux selected ($\rm{F_{\rm{3.6\,\mu m}}} > 10\,\rm{\mu Jy}$) sample in the Bo\"{o}tes field based on the multi-wavelength catalogues by \citet{Kondapally2021}. \citet{Smith2021} performed SED fitting for a similar mid-infrared sample in the ELAIS-N1 field (one of the other LoTSS-Deep fields) using \textsc{magphys} \citep{daCunha2008} to derive galaxy properties. This was then subsequently repeated for the other LoTSS-Deep fields, including the Bo\"{o}tes field out to $z \sim 1.5$ (Smith et al. priv. communication). Using this sample, we identify 118,193 SFGs (selected using the same  criterion as described in Sect.~\ref{sec:sclass_agn}) across $0.3 < z \leq 1.5$. We limit our analysis for this sample to $z < 1.5$ as we expect the photometric redshifts to be most-robust for galaxy-dominated systems below this redshift \citep{Duncan2021}. The resulting stacked properties for the star-forming galaxies sample are shown as green diamonds in Fig.~\ref{fig:lx_bhar_mass}. As done for the radio-detected AGN sample, we also subtract the XRB contribution to the X-ray luminosity for the SFG sample (see Sect.~\ref{sec:xrb_corr}). We discuss the interpretation of our results from Figs.~\ref{fig:lx_bhar_l150} and~\ref{fig:lx_bhar_mass} in more detail in Sect.~\ref{sec:discussion}.

\section{The nature of LERGs and HERGs}\label{sec:discussion}
In the contemporary picture of radio-loud AGN, we expect the LERGs and HERGs to trace the radiatively inefficient and radiatively efficient AGN, respectively. Observations at low-redshift, in particular, show that LERGs typically accrete at $\lesssim$ 1 per cent of the Eddington-scaled accretion rates \citep[e.g.][]{Hardcastle2007,2012MNRAS.421.1569B,Mingo2014}; this characteristic accretion rate threshold is motivated by the well-studied state transition observed in X-ray binaries. However, more recent deep radio surveys using both photometric and spectroscopic samples find a considerable overlap in the accretion rates of LERGs and HERGs \citep{Whittam2022,Kondapally2025,Arnaudova2025}, including a population of SF-LERGs, as identified by \citet{Kondapally2022}, which appear to show accretion rates beyond the 1 per cent threshold \citep{Kondapally2025,Arnaudova2025,Arnaudova2026}. Adding to the ambiguity are the significant uncertainties associated with deriving black hole masses (and hence Eddington-scaled accretion rates) in most of these studies, which can broaden the \textit{intrinsic} accretion rate distributions (see discussion in \citealt{Kondapally2025}). Moreover, while \citet{Kondapally2025} and \citet{Arnaudova2025} find that the LERGs at the higher end of the accretion rate distribution are hosted within star-forming galaxies (i.e. the SF-LERGs studied here), uncertainties and potential contamination from star-formation processes to the bolometric luminosities derived from photometry and from spectra (via the [O\,III] line) can complicate the interpretation. 

In this context, our results in Figs.~\ref{fig:lx_bhar_l150} and~\ref{fig:lx_bhar_mass} highlight that there is a clear difference in the \textit{average} X-ray luminosities of both the HERGs and RQ-AGN compared to the LERG population (both Q-LERGs and SF-LERGs). The HERGs and RQ-AGN both show significantly higher average X-ray luminosities (typically by an order of magnitude or more) across all radio luminosities and redshifts studied. These results suggest that the LERGs -- regardless of their star-formation activity -- likely trace a distinct population of radio-AGN, with significantly lower accretion rates than the HERGs or RQ-AGN (see Fig.~\ref{fig:lx_bhar_l150}; bottom panel). We can translate our $\lambda_{sL_{X}}$ to $\lambda_{\rm{Edd}}$ using a typical bolometric correction factor $k_{\rm{bol}} =$ 20 \citep{Lusso2012,Aird2015} and a scaling relationship between black hole mass and stellar mass of $M_{\rm{BH}} \approx 0.002 M_{\star}$ \citep{Haring2004}; these conversions carry considerable uncertainties and scatter but can be informative for comparison. Using these relations, in Figs.~\ref{fig:lx_bhar_l150} and~\ref{fig:lx_bhar_mass}, we show the specific X-ray luminosities corresponding to $\lambda_{\rm{Edd}} = $1\% and 0.01\% as grey lines for illustrative purposes. Under these assumptions, the Q-LERGs typically show $\lambda_{\rm{Edd}} \sim 0.01-0.1\%$ across all redshifts. In contrast, our SF-LERGs have typical $\lambda_{\rm{Edd}} \sim 0.1\%$ at $z \sim 0.3$, which increases with redshift, approaching $\lambda_{\rm{Edd}} \sim 1\%$ at $z \sim 2$. This increase in average specific X-ray luminosities (and hence $\lambda_{\rm{Edd}}$) with redshift for the SF-LERGs suggests that the overlap in the \textit{overall} population-level accretion rates of LERGs and HERGs found by recent photometric and spectroscopic studies \citep[e.g.][]{Whittam2022,Kondapally2022,Arnaudova2025} may be due to the SF-LERGs accreting, on average, at higher accretion rates at higher redshifts. 


Fig.~\ref{fig:lx_bhar_l150} shows that the average X-ray luminosities of HERGs and RQ-AGN in a given redshift bin show only a modest increase with increasing radio luminosity. We note that in our intermediate redshift bins ($0.5 < z < 1.5$), the average X-ray luminosities for the HERGs and RQ-AGN typically increase by a factor of $\lesssim$ 3-5 for an increase of over 2\,dex in radio luminosity. The stacked X-ray luminosities of LERGs show an even weaker dependence on radio luminosity. These results suggest that the radio emission, which traces the synchrotron emission from the jets, may not be coupled with the X-ray emission, which is typically associated from the accretion disc or the base of the jet \citep[see also][]{Delvecchio2018,Delvecchio2022,Pennock2025}. In Fig.~\ref{fig:lx_bhar_l150}, we also find that the average X-ray luminosities (and hence the $\lambda_{sL_{X}}$) for each AGN class (except for the Q-LERGs) increases with increasing redshift at a given radio luminosity; these results are qualitatively consistent with \citet{Delvecchio2018,Delvecchio2022} and suggest that radio-detected AGN are tracing more enhanced accretion activity at higher redshifts. It is plausible that the increase in molecular gas fractions with redshift \citep[e.g.][]{Tacconi2018}, which lead to more efficient star-formation, may also contribute to the increase in the average accretion rates of the black hole \citep[e.g.][]{Stanley2015}. We note that while the SF-LERG population also shows an increase in their average $\lambda_{sL_{X}}$ at higher redshifts, however this AGN activity is occurring at much lower accretion rates compared to the HERGs or RQ-AGN at fixed stellar mass and radio luminosity. This may indicate that although the hosts of SF-LERGs may on average have more abundant supply of cold gas, this gas does not appear to reach the nuclear region surrounding the black hole to result in a significant change in its accretion properties. The physical processes governing the increase in accretion rates and how they may be different from the HERGs remains unclear at present. In contrast, the lack of a significant redshift evolution in the specific X-ray luminosities for the Q-LERGs in Fig.~\ref{fig:lx_bhar_l150} compared to the other AGN classes is consistent with these sources likely being fuelled by hot gas \citep[e.g.][]{Best2006,Hardcastle2007}, representing the low accretion rate tail of the LERG population \citep[see also][]{Kondapally2025}. 

Our results in Fig.~\ref{fig:lx_bhar_mass} show that the average $L_{X,\rm{2-10\,keV}}$ and $\lambda_{sL_{X}}$ typically increases with increasing stellar mass for the SFG sample, and also increases with increasing redshift. These results are in good agreement with other measurements from the literature of the so-called `AGN main sequence', where more massive galaxies trace more efficient black hole activity, which also increases with redshift \citep[e.g.][]{Mullaney2012,Aird2018,Yang2017,Carraro2020,Ito2022}. Although we note that our data typically probes more massive host galaxies than most previous X-ray selected studies. Our results are also consistent with the recent study by \citet{Guetzoyan2025} who used an optical-selected galaxy sample in the Bo\"{o}tes field to study the Eddington-scaled black hole accretion rates as a function of stellar mass (if we adopt the same X-ray to bolometric luminosity correction and stellar-mass to black hole mass scaling relation as in their work). 

Comparing the results for the different AGN populations in our study, we find that as a function of stellar mass, the average $L_{X,\rm{2-10\,keV}}$ and $\lambda_{sL_{X}}$ both increase with increasing redshift for the HERGs, RQ-AGN, and the SF-LERGs. This behaviour is qualitatively similar to the broader SFG population, providing additional evidence in support of the radio-detected AGN undergoing more efficient accretion at higher redshifts (for a fixed stellar mass). Whereas for the Q-LERGs, whose number density declines rapidly at $z \gtrsim 1$ \citep{Kondapally2022}, our results appear to show a mild evolution with redshift but at $z > 1$, these results are mostly based on upper limits. In a given redshift bin, for all four of the radio-detected AGN samples, we find that the average $L_{X,\rm{2-10\,keV}}$ shows little-to-no correlation with stellar mass. Such a flat dependence of $L_{X,\rm{2-10\,keV}}$ on stellar mass for the AGN samples may indicate that the radio-selected AGN at higher masses may be accreting at lower rates (as seen in Fig.~\ref{fig:lx_bhar_mass}; bottom panel), resulting in lower than expected X-ray luminosities.

\section{Conclusions}\label{sec:conclusions}
In this paper, we have used data from the LoTSS Deep Fields survey and the \textit{Chandra} Deep Wide-field Survey to study the average X-ray properties of four different classes of radio-detected AGN across $\sim$9.5\,deg$^{2}$ of the Bo\"{o}tes field. The deep, wide-area optical to far-infrared imaging available in this field enables detailed source classification, AGN identification, and robust galaxy properties to be determined. Using this information, we constructed a sample of 2840 radio-detected AGN that are complete in radio-luminosity within a given redshift bin across $0.3 < z \leq 2$. This was further split into 298 HERGs, 839 RQ-AGN, 659 Q-LERGs, and 1044 SF-LERGs. We combined the radio-AGN with deep \textit{Chandra} imaging to perform X-ray stacking to study how the average X-ray luminosities ($L_{X,\rm{2-10\,keV}}$) and the average specific X-ray luminosities ($\lambda_{sL_{X}}$; average X-ray luminosity scaled by the stellar mass, which we use as a proxy for the accretion rates) depend on the four radio-AGN classes as a function of radio luminosity, stellar mass, and redshift. We also compared our results to both radio and non-radio detected studies in the literature. Our main conclusions are as follows:

\begin{enumerate}
    \item The LERGs (regardless of their host-galaxy star-formation activity) show significantly lower (by an order of magnitude) X-ray luminosities compared to the HERGs and RQ-AGN.
    \item The average $L_{X,\rm{2-10\,keV}}$ for all four AGN classes shows no correlation with $L_{\rm{150\,MHz}}$ within a given redshift bin, but the $L_{X,\rm{2-10\,keV}}$ typically increases with increasing redshift.
    \item The average $\lambda_{sL_{X}}$ for the HERGs, RQ-AGN, and SF-LEGRs increases by a factor of 10 from $z \sim 0.3$ to $z \sim 2$; this indicates that the radio-detected AGN in general are undergoing more enhanced accretion at higher redshifts.
    \item The average $L_{X,\rm{2-10\,keV}}$ for all four AGN classes shows a flat dependence on stellar mass. We suggest that this may be due to radio-AGN in higher mass galaxies accreting at lower accretion rates.
\end{enumerate}

In summary, our results suggest that the LERGs (hosted in both star-forming and quiescent galaxies) may represent a distinct population of AGN, which are accreting at much lower rates compared to the HERGs and RQ-AGN; these trends remain as a function of radio luminosity and stellar mass, across redshift. The high X-ray luminosities and $\lambda_{sL_{X}}$ found for HERGs and RQ-AGN, and their increase with redshift, suggests that the level of accretion may be linked to the availability of cold gas, which increases at higher redshifts, resulting in more efficient accretion. The SF-LERGs also show such an increase in their $\lambda_{sL_{X}}$ with redshift albeit still at much lower accretion rates than HERGs and RQ-AGN. This may suggest that while the increasing availability of cold gas may fuel star-formation, it may not lead to significantly enhanced radiatively-efficient accretion in SF-LERGs. 

\section*{Acknowledgements}

RK acknowledges support from the Leverhulme Trust through a Leverhulme Early Career Fellowship. TH acknowledges funding from the Public Scholarship, Development, Disability and Maintenance Fund of the Republic of Slovenia. JA acknowledges support from a UKRI Future Leaders Fellowship (grant code: MR/Y019539/1). PNB acknowledges support from the UK Science and Technologies Facilities Council (STFC) via grant ST/Y000951/1. KJD acknowledges support from STFC through an Ernest Rutherford Fellowship (grant number ST/W003120/1). LKM is grateful for support from a Future Leaders Fellowship from UKRI [MR/Y020405/1] and LOFAR-UK via STFC [ST/V002406/1]. SS and DJBS acknowledge support from the UK STFC via the grant ST/X508408/1. SD acknowledges support from the Leverhulme Trust via Research Project Grant RPG-2025-078. We thank Ryan Hickox for helpful discussions on the results in the paper.

This paper is based (in part) on data obtained with the International LOFAR Telescope (ILT) under project codes LC0\_015, LC2\_024, LC2\_038, LC3\_008, LC4\_008, LC4\_034 and LT10\_01. LOFAR \citep{2013A&A...556A...2V} is the Low Frequency Array designed and constructed by ASTRON. It has observing, data processing, and data storage facilities in several countries, which are owned by various parties (each with their own funding sources), and which are collectively operated by the ILT foundation under a joint scientific policy. The ILT resources have benefitted from the following recent major funding sources: CNRS-INSU, Observatoire de Paris and Université d'Orléans, France; BMBF, MIWF-NRW, MPG, Germany; Science Foundation Ireland (SFI), Department of Business, Enterprise and Innovation (DBEI), Ireland; NWO, The Netherlands; The Science and Technology Facilities Council, UK; Ministry of Science and Higher Education, Poland. For the purpose of open access, the author has applied a Creative Commons Attribution (CC BY) licence to any Author Accepted Manuscript version arising from this submission. This research made use of {\sc Astropy}, a community-developed core Python package for astronomy \citep{astropy:2013, astropy:2018} hosted at \url{http://www.astropy.org/}, and of {\sc Matplotlib} \citep{hunter2007matplotlib}.

\section*{Data Availability}

The dataset used in this study comes primarily from the LoTSS Deep Fields Data Release 1. The corresponding radio data are presented by \citet{Tasse2021}, the multi-wavelength data, host-galaxy counterparts, and photometric redshifts are presented by \citet{Kondapally2021} and \citet{Duncan2021}, with the source classifications presented by \citet{Best2023}. The images and catalogues are publicly available at \url{https://lofar-surveys.org/deepfields.html}. The SED fitting output for the mid-infrared parent sample is presented by \citet{Smith2021}. The X-ray data used comes from the \textit{Chandra} Deep Wide-Field Survey (CDWFS) presented by \citet{Masini2020}.




\bibliographystyle{mnras}
\bibliography{lofar_deep} 








\bsp	
\label{lastpage}
\end{document}